\makeatletter
\def\input@path{{../template}}
\makeatother

\documentclass[summary]{ursi}
\usepackage{booktabs}
\usepackage{subfig}
\usepackage{xcolor}
\usepackage{paralist}
\usepackage{tikz}

\usepackage[absolute,overlay]{textpos}
\usepackage[yyyymmdd,hhmmss]{datetime}

\title{Measurement of Liquid Water Content in Snow from Density-Insensitive Microwave Attenuation}

\author{Marco Niederberger*\affref{ost}, Sebastian Droz\affref{ost},
  Shaarujan Kamalanathan\affref{ost}, \\ Michel A. Nyffenegger\affref{ost}, Albert Loichinger\affref{ost}, and Hans-Dieter Lang\affref{ost}}

\affiliation{%
  \aff{ost}{OST Eastern Switzerland University of Applied Sciences, Rapperswil, SG, Switzerland, http://www.ost.ch}
}

\newcommand{\kgm}{\text{kg/m\textsuperscript{3}}}
\newcommand{\figref}[1]{Fig.~\ref{#1}}
\begin{document}

\maketitle

\begin{abstract}
    A measurement principle for determining liquid water content (LWC) in snow is presented that does not require a priori knowledge or separate measurement of snow density.
    LWC affects the imaginary part of the effective permittivity of snow, which is linked to microwave attenuation along a microstrip transmission line embedded in the snowpack, independent of snow density.
    By analyzing the attenuation over a range of microwave frequencies, the corresponding LWC can be inferred.
    Full-wave simulations indicate a sensitivity of approximately 2 dB per \% LWC, while experimental results confirm the insensitivity to snow density, highlighting the potential for remote snow characterization.
\end{abstract}

\section{Introduction}
Predicting wet snow avalanches remains a challenging task due to limited knowledge of the snowpack and snow microstructure, and a lack of real-time field measurements.
Among snow properties, liquid water content (LWC) plays a key role in governing mechanical strength and avalanche formation \cite{wetSnowAvalanche, toolWet}.
Even small amounts of liquid water can significantly impact snowpack stability, making accurate monitoring of LWC essential for reliable avalanche prediction.
For wet-snow avalanche studies, snow is commonly classified into four regimes based on LWC \cite{snowClassification}:

\begin{compactitem}
    \item \emph{dry snow} with 0\,\% LWC.
    \item \emph{moist snow} with LWC below 3\%
    \item \emph{wet snow} with LWC between 3\% and 8\%
    \item \emph{very wet snow} with LWC above 8\%.
\end{compactitem}

Snow in the \emph{wet} and \emph{very wet} regimes exhibits substantially reduced mechanical strength and is therefore particularly relevant for avalanche prediction \cite{wetSnowAvalanche}.

This work introduces a new electromagnetic sensing principle for real-time monitoring of LWC in snow.
The approach leverages the dependence of snow's complex relative permittivity on water content, where the imaginary part increases with LWC independently of snow density.
A microstrip transmission line embedded in the snowpack enables estimation of LWC from the corresponding attenuation of microwave signals traveling along the line.
By analyzing the attenuation over a range of microwave frequencies, the effective permittivity of the surrounding snow can be inferred and the LWC can be determined.
The method is suitable for continuous, in-situ measurements and deployment in remote or difficult-to-access locations.

\section{Related Work}
Current methods for determining LWC in snow are based on capacitance measurements, time-domain reflectometry (TDR), GNSS-based methods, or vector network analyzer (VNA) measurements in the far-field.
Each approach has inherent limitations related to the underlying measurement principle.

The Denoth meter~\cite{Denoth1994, fundamentals:denothSLF} essentially consists of a coplanar capacitor on a PCB and an external quartz crystal, together acting as frequency-determining elements within an oscillator circuit.
The electric permittivity of the snow modifies the capacitance, resulting in a measurable frequency shift.
This approach is only sensitive to the real part of the permittivity, requiring the dry snow density to determine the LWC.
Knowledge of the dry snow density is not available for in-situ measurements.

Waldner et al.~\cite{Waldner2001} and Schneebeli et al.~\cite{Schneebeli1998} used time-domain reflectometry (TDR) to estimate snow permittivity by analyzing signal reflections from an aluminum-rod sensor.
While this method enables in-situ measurements, it again primarily captures the real part of the relative permittivity of snow.
Consequently, LWC can be determined only when combined with separate density measurements.

Koch et al.~\cite{Koch2014} demonstrated using low-cost GPS receivers to estimate the LWC in snow by comparing the signal-to-noise ratios of two receivers placed above and below the snowpack.
Although this approach does not require a separate density measurement, it relies on two spatially separated devices, one of which must remain exposed above the snow surface.
Furthermore, this method only provides a bulk measurement of the entire snowpack and therefore provides limited spatial resolution.

Finally, VNA-based measurement techniques have also been reported \cite{vnabased}.
These typically require extracting a snow sample and placing it between two horn antennas under controlled conditions, making them unsuitable for automated, continuous, or remote LWC monitoring in the field.

\section{Methodology}

\subsection{Material Characterization}

Dry snow consists solely of ice and air.
As the ice begins to melt or due to rainfall, liquid water becomes present within the snowpack. This is quantified by the liquid water content (LWC) in snow, defined as the ratio of the liquid water volume $V_W$ to the volume of snow $V_S$ \cite{snowClassification}:
\begin{equation}
    \theta = \dfrac{V_W}{V_S}\,.
\end{equation}
LWC is independent of snow density: two snow volumes $V_S$ of equal size containing the same water volume $V_W$ exhibit identical LWC $\theta$, regardless of the density of the snow.

\begin{figure}[!b]
    \centering
    \vspace*{-3mm}
    \includegraphics[width=0.96\linewidth, clip, trim=0 2mm 0 2.5mm]{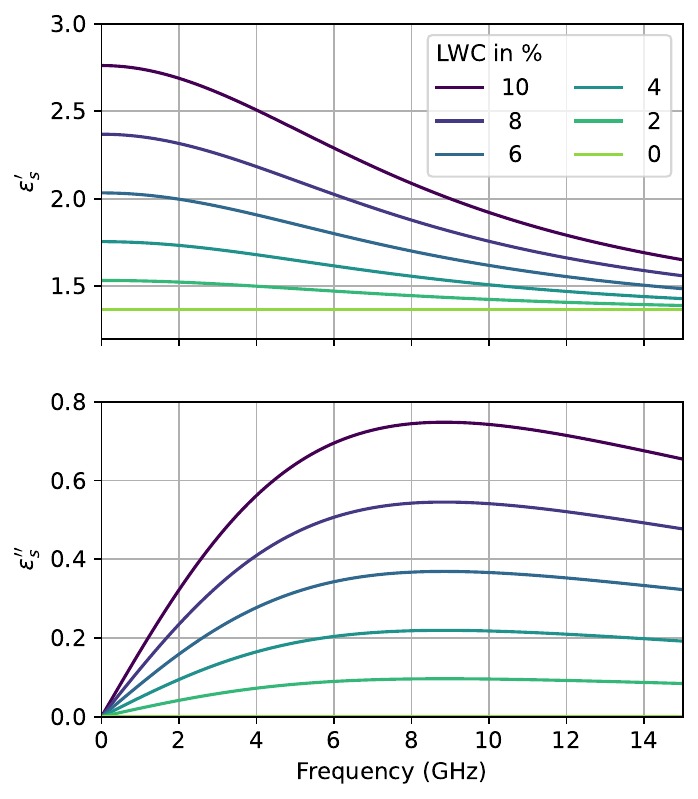}
    \caption{Relative permittivity $\varepsilon_s = \varepsilon_s^\prime - j \varepsilon_s^{\prime\prime}$ of snow for different LWC (density $\rho_s = 450$ kg/m\textsuperscript{3}).}
    \label{fig:taylor}
\end{figure}

The complex relative permittivity of snow
\begin{subequations}
    \label{eq:taylor:all}
    \begin{equation}
        \varepsilon_{r,s} = \varepsilon'_{r,s} - j\varepsilon''_{r,s}
    \end{equation}
    can be modeled using \cite{complexDielectricSnow}
    \begin{align}
        \varepsilon'_{r,s}  & = k\,\varepsilon'_{r,w}(f) + 1 + 1.7(\tilde{\rho}_s-\theta) + 0.7(\tilde{\rho}_s-\theta)^2, \\
        \varepsilon''_{r,s} & = k\,\varepsilon''_{r,w}(f), \label{eq:taylor}                                              \\
        k                   & =0.1\theta+0.8\theta^2
    \end{align}
\end{subequations}
where $\theta$ denotes the LWC ranging from 0 to 10\,\%, and $k$ is an empirical quadratic correction factor.
Moreover, $\varepsilon_{r,w}'(f)$ and $\varepsilon_{r,w}''(f)$ are the real and imaginary parts, respectively, of the relative permittivity of water at the frequency of operation $f$,
and $\tilde{\rho}_s = \rho_s/\rho_w$ denotes the density of dry snow relative to the density of water.
The resulting behavior of the real and imaginary parts of $\varepsilon_{r,s}$ is illustrated in \figref{fig:taylor}.

The relative permittivity of water at the frequency $f$ can be modeled using a Debye relaxation model \cite{complexDielectricSnow}
\begin{multline}
    \varepsilon_{r,w}(f)
    =
    \varepsilon'_{r,w}(f) - j\varepsilon''_{r,w}(f) \\[-1.5mm]
    =
    \varepsilon_{r,w}(\infty) + \frac{\varepsilon_{r,w}(0) - \varepsilon_{r,w}(\infty)} {1 + j\,2\pi f \tau},
\end{multline}
where $\varepsilon_{r,w}(0) \approx 88$ is the relative permittivity of water at DC ($f=0$),
$\varepsilon_{r,w}(\infty) \approx 4.9$ the corresponding high-frequency limit ($f \to \infty$), and
$\tau\approx17\,\text{ps}$ its relaxation time.

\subsection{LWC Measurement Principle}
According to \eqref{eq:taylor}, the imaginary part of the relative permittivity of snow $\varepsilon''_{r,s}$ depends solely on the LWC.
This property forms the basis of the proposed measurement principle, which exploits how a guided electromagnetic wave in a microstrip transmission line embedded in snow interacts with the surrounding dielectric materials \cite{lossyDielectric}.
As LWC increases, $\varepsilon_{r,s}''$ rises independently of snow density, causing greater power dissipation and a measurable increase in attenuation along the microstrip line.
This can be captured using amplitude-only measurements with a simple power detector, without requiring phase information.

The real part of the relative permittivity \(\varepsilon'_{r,s}\) exhibits some dependence on snow density in addition to LWC, primarily affecting the wave impedance and phase constant. However, its influence on attenuation is minor.
To further increase robustness, attenuation is therefore evaluated over a range of microwave frequencies, e.g., 5-7\,GHz, rather than at a single frequency.
Consequently, attenuation measurements enable simple estimation of LWC in snow without a priori knowledge of snow density.

\subsection{LWC Sensor Proof-of-Concept Design}
To evaluate the proposed concept, a sensor was designed to measure LWC ranging from 0\,\% to 10\,\%, covering the entire LWC range relevant for wet-snow avalanche studies. PTFE (ZYF255DA) was selected as substrate material of the microstrip line due to its low permittivity ($\varepsilon_r = 2.55$) and hydrophobic properties, making it well suited for operation in wet snow.

\begin{figure}[!htb]\centering
    \includegraphics[width=0.9\linewidth, trim=60 150 180 40, clip]{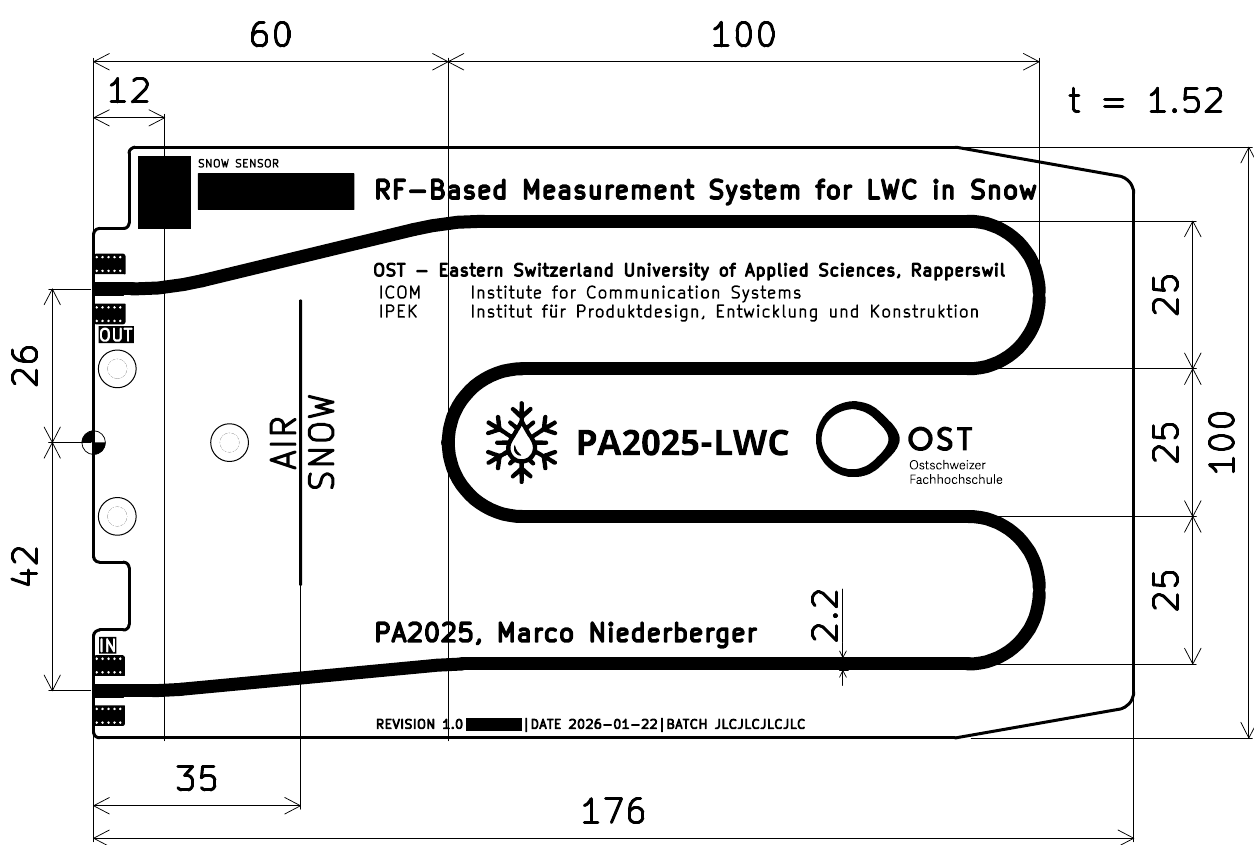}
    \caption{Geometry of the designed snow sensor. All dimensions are in millimeters.}
    \label{fig:sensorKiCAD}
    \vspace*{-3mm}
\end{figure}

The sensor design was developed and simulated using Ansys HFSS.
Snow was modeled as an effective homogeneous dielectric layer with complex relative permittivity, as defined in \eqref{eq:taylor:all}.
Considering the limitations of the involved measurement equipment, the microstrip length and width were designed to obtain a sensitivity of at least 2\,dB change in attenuation per \% LWC, while keeping the maximum attenuation below 50\,dB.
For the aforementioned substrate, this results in a strip length of 560\,mm and a width of 2.2\,mm for 1.52\,mm substrate thickness and 35\,\textmu{}m copper cladding.
The sensor geometry and dimensions are illustrated in \figref{fig:sensorKiCAD}.

\begin{figure}[!b]\centering
    \vspace*{-4mm}
    \includegraphics[width=0.95\linewidth]{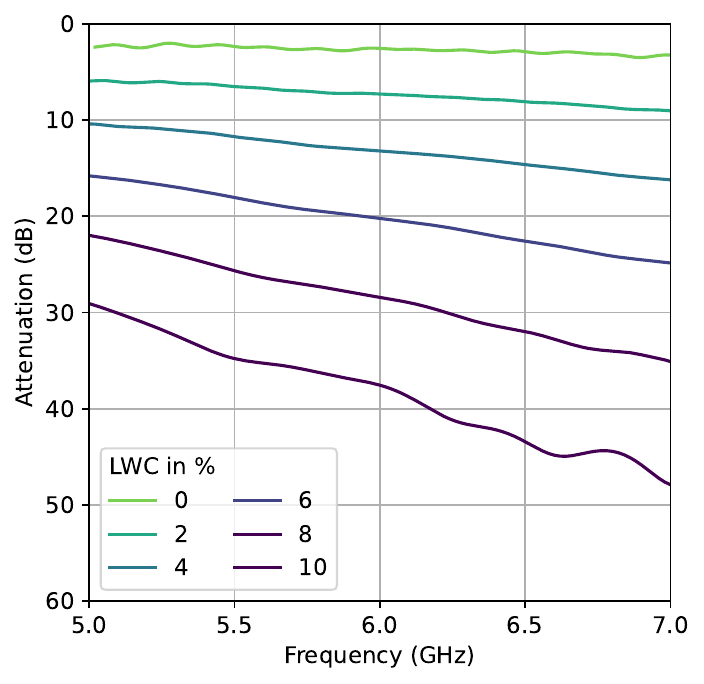}
    \vspace*{-2mm}
    \caption{Simulated attenuation for a $50\,\Omega$-microstrip line of 560\,mm length on PTFE substrate and various LWC of snow and a density $\rho_s = 450$ kg/m\textsuperscript{3}.}
    \label{fig:lwc_dependency}
    \vspace*{-1mm}
\end{figure}

\figref{fig:lwc_dependency} shows the simulated attenuation over the considered LWC range.
At 5\,GHz, attenuation increases from approximately 2.5\,dB for \textit{dry snow} to around 30\,dB for \textit{very wet snow} with an LWC of 10\,\%.
Within the \emph{moist snow} regime, the sensor exhibits a sensitivity of roughly 2\,dB per \% change in LWC.

\subsection{Experimental Verification in the Field}
\begin{figure}[ht]
    \centering
    \begin{tikzpicture}
        \node[anchor=south west, inner sep=0] (image) at (0,0)
        {\includegraphics[width=\linewidth,clip,trim={80mm 0 22mm 0}]{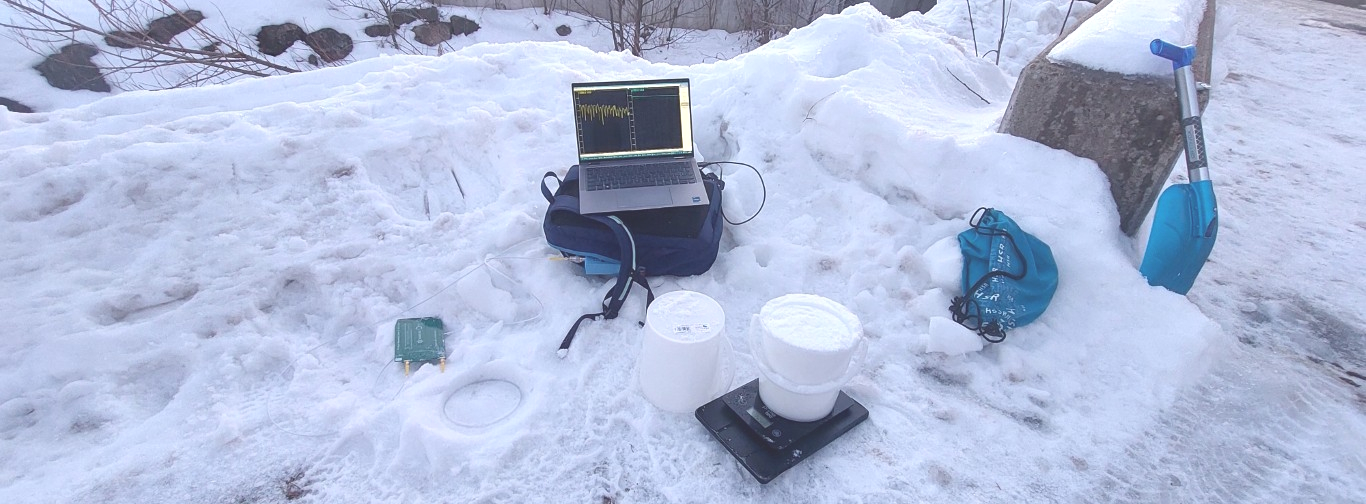}};
        \begin{scope}[x={(image.south east)}, y={(image.north west)}]

            \tikzset{
                lab/.style={
                        font=\sffamily\fontsize{10}{12}\selectfont,
                        text=black,
                        fill=white,
                        fill opacity=0.8,
                        text opacity=1,
                        rounded corners=2pt,
                        inner sep=3pt
                    }
            }
            \tikzset{
                line/.style={
                        line width=0.3mm,
                        line cap=round
                    }
            }

            \draw[line] (0.15,0.18) -- (0.18,0.25);
            \node[lab, anchor=north] at (0.15,0.18) {LWC sensor};

            \draw[line] (0.25,0.5) -- (0.33,0.48);
            \node[lab, anchor=east] at (0.25,0.5) {VNA};

            \draw[line] (0.61,0.53) -- (0.54, 0.36);
            \node[lab, anchor=west] at (0.61,0.53) {Snow sample};

            \draw[line] (0.61,0.12) -- (0.55, 0.15);
            \node[lab, anchor=west] at (0.61,0.12) {Scale};

        \end{scope}
    \end{tikzpicture}
    \caption{Measurement setup in the field with the snow sample, the LWC sensor and the VNA.}
    \label{fig:setup}
\end{figure}
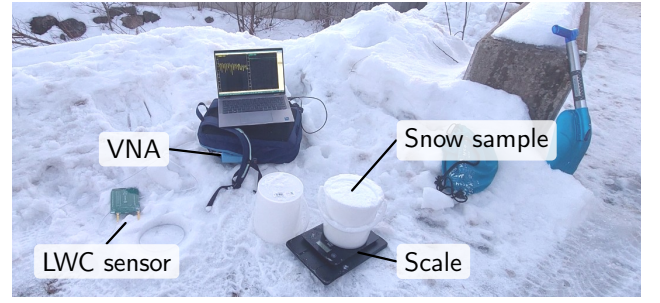

Field experiments were conducted to verify the density independence of the proposed measurement principle, which constitutes its primary advantage over existing methods.
For this purpose, dry snow samples were collected in the field and measured both before and after mechanical compaction, as shown in \figref{fig:setup}.
The absence of liquid water was ensured by ambient temperatures as low as --\hspace*{1pt}4\,°C during sampling.
Snow density was determined from the measured mass and volume of each sample.
If the measurement principle indeed proves to be independent of snow density, the resulting attenuation curves are expected to remain closely aligned despite significant density variations.

\section{Results}

\begin{figure}[t!]
    \centering
    \subfloat[][]{\includegraphics[width=0.95\linewidth, trim=0 315 0 0, clip]{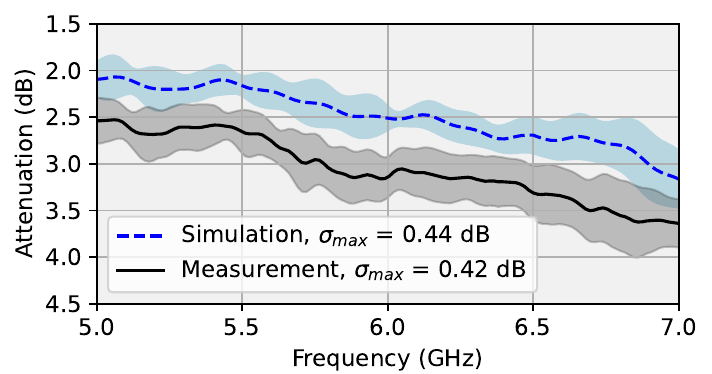}}

    \subfloat[][]{\includegraphics[width=0.95\linewidth, trim=0 0 0 182, clip]{densities_ROT_both.pdf}}
    \vspace*{-2mm}

    \caption{Attenuation measurements in dry and wet snow samples:
        (a) Zoomed-in view of the dry-snow measurements shown in (b), illustrating the mean value and standard deviation of simulated and measured attenuation for snow densities between 200\,\kgm{} and 450\,\kgm{}.
        (b) Measured and simulated attenuation for dry and wet snow samples, where the dry-snow measurements essentially collapse into a single curve. Comparison with the wet-snow measurements highlights the negligible influence of snow density relative to the effect of LWC, consistent with the proposed measurement principle.}
    \vspace*{-2mm}
    \label{fig:density}
\end{figure}

The proposed measurement principle's independence from snow density was experimentally verified using dry snow samples with densities ranging from 200\,\kgm{} up to 450\,\kgm{}.
The corresponding simulated and measured attenuation results are shown in \figref{fig:density}(a), a zoomed-in region of (b).
For both simulation and experiment, the observed standard deviation is approximately 0.4\,dB, which, given the measured sensitivity, corresponds to an uncertainty of about 0.2\,\%~LWC.
This small variation confirms weak sensitivity to snow density; hence the measurement can be considered effectively density independent.

\figref{fig:density}(b) compares the same dry-snow data with measurements obtained from wet snow.
The measured trends are in good agreement with the simulations, although an exact absolute match would require detailed calibration using well-controlled reference samples.
More importantly, attenuation variations caused by density changes are negligible compared to the substantially larger attenuation changes resulting from variations in LWC.

In \figref{fig:density}(a), a small systematic offset of approx.~0.5\,dB is observed between simulated and measured attenuation.
The origin of this offset is not fully resolved.
However, as can be seen in \figref{fig:density}(b), it is negligible compared to the full attenuation range
and can be compensated by calibration.

\section{Summary and Conclusion}

This work presented an RF-based measurement principle for determining the liquid water content (LWC) in snow using a microstrip transmission line embedded in the snowpack.
By evaluating the frequency-dependent attenuation over a range of microwave frequencies, the LWC can be inferred from the imaginary part of the effective relative permittivity of snow.

Full-wave simulations and preliminary experimental verification demonstrate a sensitivity of at least 2\,dB per \%~LWC, while exhibiting only weak sensitivity to snow density.
For dry snow densities between 200\,\kgm{} and 450\,\kgm{}, the inferred LWC shows a standard deviation of approximately 0.2\,\% LWC, confirming the effective density independence of the proposed approach.
This represents a clear advantage over established techniques such as the Denoth meter or TDR methods, which require additional density measurements.
Such low-complexity microstrip-based sensors rely solely on amplitude measurements, require no auxiliary sensors, and are mechanically robust, making them well suited for remote deployment in snow-covered or avalanche-prone environments.

Accurate quantitative LWC estimation over the entire desired range from 0\,\% to 10\,\% will require a more extensive calibration campaign covering a wide range of snow conditions.
The development of such a calibration procedure, which is experimentally challenging due to the difficulty of producing well-controlled snow samples \cite{fundamentals:denothSLF, Schneebeli1998},
will be addressed in future work.

\section*{Acknowledgment}
The authors thank Dr. Alec van Herwijnen and the Institute for Snow and Avalanche Research (SLF) for their support.


\begin{thebibliography}{99}
    \bibitem{wetSnowAvalanche}
    I.~Reiweger, M.~Zöchling, M.~Forster, T.~Wiesinger, and C.~Mitterer, ``Wet-Snow Fracture Propagation,'' in {\em Proceedings of the International Snow Science Workshop (ISSW) 2016}, Oct. 2016.

    \bibitem{toolWet}
    C.~Mitterer, F.~Techel, C.~Fierz, and J.~Schweizer, ``{An Operational Supporting Tool for Assessing Wet-Snow Avalanche Danger},'' in {\em Proceedings of the International Snow Science Workshop (ISSW)}, 2013.

    \bibitem{snowClassification}
    C.~Fierz, R.~L. Armstrong, Y.~Durand, P.~Etchevers, E.~Greene, D.~M. McClung, K.~Nishimura, P.~K. Satyawali, and S.~A. Sokratov, {\em {The International Classification for Seasonal Snow on the Ground}}.
    \newblock IHP-VII Tech. Doc. in Hydrology N°83, IACS Contribution N°1, Paris, France: UNESCO-IHP, 2009.

    \bibitem{Denoth1994}
    A.~Denoth, ``{An Electronic Device for Long-Term Snow Wetness Recording},''
    {\em Annals of Glaciology}, vol.~19, pp.~104--106, 1994.

    \bibitem{fundamentals:denothSLF}
    F.~Wolfsperger, M.~Geisser, S.~Ziegler, and H.~Löwe,
    ``A New Handheld Capacitive Sensor to Measure Snow Density and Liquid Water Content,'' in {\em Proc. Int. Snow Sci. Work.}, (Bend, Oregon), Oct. 2023.

    \bibitem{Waldner2001}
    P.~Waldner, C.~Huebner, M.~Schneebeli, A.~Brandelik, and F.~Rau, ``{Continuous Measurements of Liquid Water Content and Density in Snow Using TDR},'' in {\em Proc. of the 2nd Int. Symp. and Workshop on TDR for Innovative Geotech. Appl. (TDR 2001)}, Sept. 2001.

    \bibitem{Schneebeli1998}
    M.~Schneebeli, C.~Coléou, F.~Touvier, and B.~Lesaffre, ``{Measurement of
    Density and Wetness in Snow using Time-Domain Reflectometry},'' {\em Annals
    of Glaciology}, vol.~26, pp.~69--72, 1998.

    \bibitem{Koch2014}
    F.~Koch, M.~Prasch, L.~Schmid, J.~Schweizer, and W.~Mauser, ``{Measuring Snow
    Liquid Water Content with Low-Cost GPS Receivers},'' {\em Sensors}, vol.~14,
    pp.~20975--20999, Nov. 2014.

    \bibitem{vnabased}
    C.~Bermond, P.~Artillan, and M.~Gay, ``{A Microwave Frequency Range Experiment for the Measurement of Snow Density and Liquid Water Content},'' {\em {IEEE} J. Sel. Topics Appl. Earth Observ. Remote Sens}, vol.~14, pp.~11197--11203, 2021.

    \bibitem{complexDielectricSnow}
    M.~Tiuri, A.~Sihvola, E.~Nyfors, and M.~Hallikaiken, ``{The Complex Dielectric
    Constant of Snow at Microwave Frequencies},'' {\em IEEE Journal of Oceanic
    Engineering}, vol.~9, no.~5, pp.~377--382, Dec. 1984.

    \bibitem{lossyDielectric}
    I.~J. Bahl and S.~S. Stuchly,
    ``{Analysis of a Microstrip Covered with a Lossy Dielectric},''
    {\em {IEEE} Trans. Microw. Theory Techn.}, vol.~28, no. 2, Feb. 1980.

\end{thebibliography}
\end{document}